\documentstyle[11pt,psfig]{article}

\makeatletter

\def\thebibliography#1{\section*{References\markboth
 {REFERENCES}{REFERENCES}}\list
  {}{\settowidth\labelwidth{0pt}\leftmargin\labelwidth    
 \advance\leftmargin\labelsep
 \usecounter{enumi}\@bibsetup}
 \def\newblock{\hskip .11em plus .33em minus -.07em}
 \sloppy\clubpenalty4000\widowpenalty4000
 \sfcode`\.=1000\relax}

\def\@bibsetup{\itemindent=-\leftmargin \itemsep=0pt
 \parsep=0pt  
 }

\def\@biblabel#1{\hfill}

\long\def\@makecaption#1#2{%
  \small
  \vskip\abovecaptionskip
  \sbox\@tempboxa{#1: #2}%
  \ifdim \wd\@tempboxa >\hsize
    #1: #2\par
  \else
    \global \@minipagefalse
    \hb@xt@\hsize{\hfil\box\@tempboxa\hfil}%
  \fi
  \normalsize
  \vskip\belowcaptionskip}

\def\bmin{b_{\rm min}}
\def\bmax{b_{\rm max}}
\def\d{{\rm d}}
\def\deriv#1#2{{\partial #1\over\partial #2}}
\def\De{D_E}
\def\Dee{D_{EE}}
\def\Nc{N_{\rm c}}
\def\pc{\,{\rm pc}}
\def\pl{p_\Lambda}
\def\phic{\phi_{\rm c}}
\def\phii{\phi_\infty}
\def\ql{q_\Lambda}
\def\rc{r_{\rm c}}
\def\rh{r_{\rm h}}
\def\rk{r_{\rm k}}
\def\rt{r_{\rm t}}
\def\rmin{r_{\rm min}}
\def\rmax{r_{\rm max}}
\def\rhoc{\rho_{\rm c}}
\def\tr{t_{\rm r}}
\def\tcc{t_{\rm cc}}
\def\trc{t_{\rm rc}}
\def\trh{t_{\rm rh}}
\def\vc{v_{\rm c}}
\def\yr{\,{\rm yr}}

\def\simless{\mathbin{\lower 3pt\hbox
   {$\rlap{\raise 5pt\hbox{$\char'074$}}\mathchar"7218$}}} 
\def\simgreat{\mathbin{\lower 3pt\hbox
   {$\rlap{\raise 5pt\hbox{$\char'076$}}\mathchar"7218$}}} 

\makeatother

\begin{document}

\title{{\bf The time-scale for core collapse in spherical star clusters}}
\author{Gerald D.\ Quinlan\\
Lick Observatory, University of California, Santa Cruz CA 95064 \\
Dept.\ of Physics and Astronomy, Rutgers University, PO Box 849, Piscataway
NJ 08855\thanks{Present address.}}
\maketitle

\begin{abstract}
The collapse time for a cluster of equal-mass stars is usually stated to be
either 330 central relaxation times ($\trc$) or 12--19 half-mass relaxation
times ($\trh$).  But the first of these times applies only to the late
stages of core collapse, and the second only to low-concentration clusters.
To clarify how the time depends on the mass distribution, the Fokker-Planck
equation is solved for the evolution of a variety of isotropic cluster
models, including King models, models with power-law density cusps of
$\rho\sim r^{-\gamma}$, and models with nuclei.  High-concentration King
models collapse faster than low-concentration models if the time is measured
in units of $\trh$, but slower if it is measured in units of $\trc$.  Models
with cusps evolve faster than King models, but not all of them collapse:
those with $0<\gamma<2$ expand because they start with a temperature
inversion.  Models with nuclei collapse or expand as the nuclei would in
isolation if their central relaxation times are short; otherwise their
evolution is more complicated.  Suggestions are made for how the results can
be applied to globular clusters, galaxies, and hypothetical clusters of dark
stars in the centers of galaxies.
\end{abstract}

\section{Introduction}

The dynamical processes responsible for core collapse are now largely
understood (Spitzer 1987).  For many star clusters the collapse can be
divided into two stages.  The first is driven by the approach to thermal
equilibrium---a state that is unreachable because of the finite escape
velocity.  The core contracts to conserve energy as stars evaporate from the
high end of the velocity distribution.  If this were all that happened the
collapse time (called the evaporation time in this context) would be
long---a hundred or more half-mass relaxation times.  But as the collapse
proceeds the core grows hotter and begins transferring energy to the cooler
surrounding stars.  This causes it to contract further and grow even
hotter---an instability known as the gravothermal catastrophe that drives
the core in a self-similar manner to zero size and infinite density, with
the time remaining until complete collapse at any instant being about 330
times the central relaxation time at that instant.  Complications arise near
the end when the core is small---stars can merge, massive stars can evolve,
binaries can form and harden, possibly stopping and reversing the collapse
(reviewed by Goodman 1989, 1993)---but most of the time needed to reach this
state is spent near the start where the evolution is simple.

Yet despite the progress that has been made in understanding this evolution,
the estimation of collapse times for real star clusters remains confusing.
References on the subject usually give two collapse times. The first, 330
central relaxation times ($\trc$), applies only to the late, self-similar
stage of core collapse.  The second, 12--19 half-mass relaxation times
($\trh$), applies only to low-concentration models like the Plummer model in
which the central and half-mass relaxation times are nearly the same.  These
are poor models for real star clusters.  Globular clusters have central
relaxation times that are typically ten times shorter than their half-mass
relaxation times, sometimes a hundred or a thousand times shorter.  Galaxies
are even more concentrated: many have densities that continue rising to the
innermost observable radius; some have dense nuclear star clusters.
Although their half-mass relaxation times are much longer than their ages,
their central relaxation times can be short.  The simple collapse times
quoted above are a poor guide to how relaxation will affect these systems
(as Heggie and Mathieu~1986 and Goodman~1993 have stressed).

To broaden the class of models for which core collapse has been studied,
this paper uses the Fokker-Planck equation to follow the evolution of a
variety of spherical cluster models, including King models (traditionally
used to fit globular clusters), a family of models with power-law density
cusps of $\rho\sim r^{-\gamma}$ (similar to the cusps observed in galaxies),
and some two-component models with nuclei.  The calculations are simplified
in many ways---the velocity distribution is assumed to be isotropic, the
stars are treated as unevolving point masses, all of them having the same
mass (except in the models with nuclei), mass loss from a tidal boundary is
ignored, binaries are ignored---because the goal is not to study accurate
models for real star clusters but to study a wide variety of idealized
models and to isolate the dependence of the collapse time on the mass
distribution.  Binaries and stellar collisions become important only during
the late stages of core collapse; the other neglected complications have
been studied by others and will not be re-examined here.

The calculations show, as expected, that the simple collapse times quoted
above are not applicable to all systems.  Low-concentration King~models do
collapse in about 12--19$\,\trh$, but this is much shorter than 330$\,\trc$
for them.  High-concentration King models collapse faster than
low-concentration models if the time is measured in units of $\trh$, but
slower if it is measured in units of $\trc$.  Models with cusps evolve
faster than King models because they start far from thermal equilibrium.
But not all of them collapse: those with $\gamma<2$ undergo a gravothermal
expansion because they start with a temperature inversion.  Models with
nuclei are more complicated: if the nucleus has a short relaxation time it
collapses or expands as it would in isolation; otherwise its evolution is
determined its interaction with the rest of the model.

The final section of the paper discusses possible applications of the
results to globular clusters, galaxies, and hypothetical clusters of dark
stars in the centers of galaxies.  One technical detail is discussed in an
appendix: the variation of the Coulomb logarithm with position.  This
changes the evolution times only slightly if the number of stars is large,
but it can be important for detailed comparisons of Fokker-Planck
calculations with N-body experiments.

\section{Computational method}

\subsection{The Fokker-Planck equation} 

The Fokker-Planck equation describes the evolution of a stellar distribution
function resulting from weak two-body encounters, which are more important
than strong encounters by a factor of order $\ln N$ when the number of stars
$N$ is large.  Although the equation can be written for any cluster geometry
and any distribution function, its practical solution requires some
simplifying assumptions to be made.  The calculations done here assume the
cluster to be spherical and the distribution function to depend only on
energy (so the velocity distribution is isotropic); the diffusion
coefficients are evaluated in the local approximation and the equation is
orbit averaged (Binney and Tremaine 1987).  It then takes the form (Spitzer
1987)
\begin{equation}                                                  \label{eq-fp}
  \deriv{f}{t} - {1\over p}\deriv{q}{t}\deriv{f}{E} = {1\over p}
  \deriv{}{E}\left( \De f + \Dee\deriv{f}{E} \right),
\end{equation}
where the coefficients $\De$ and $\Dee$ are
\begin{eqnarray}                                                  \label{eq-de}
  \De(E) &=& 16\pi^2 G^2 m^2 \ln\Lambda \int^E_{\phic}\!\! dE_1\, f_1 p_1, \\
  \Dee(E)&=& 16\pi^2 G^2 m^2 \ln\Lambda                          \label{eq-dee}
         \left( \int^E_{\phic}\!\! dE_1\, f_1 q_1 +
               q\int^{\phii}_{E}\!\! dE_1\, f_1 \right),
\end{eqnarray}
and the phase-space integrals $p$ and $q$ are
\begin{equation}                                                  \label{eq-pq}
   p(E) = \deriv{q}{E}, \qquad 
   q(E) = {1\over3}\int_0^{\phi^{-1}(E)}\!\! dr\,r^2 
                   \left[2E-2\phi(r)\right]^{3/2}.
\end{equation}
The density and potential are computed from the distribution function by
\begin{equation}                                                 \label{eq-rho}
  \nabla^2\phi(r) = 4\pi G\rho(r) = 
                    16\pi^2 G m\int^{\phii}_{\phi(r)}\!\! dE\,  
                    f \left[2E-2\phi(r)\right]^{1/2};
\end{equation}
$\phic$ and $\phii$ are the values of the potential at $r=0$ and $r=\infty$
($\phii$ is assumed to be zero in the rest of the paper).  The Coulomb
logarithm $\ln\Lambda$ is treated as a constant.  Its value---usually taken
to be $\ln(kN)$ with some numerical coefficient $k$ of order unity---need
not be specified here because it can be absorbed into the unit of time (the
relaxation time).  The changes that result when $\ln\Lambda$ varies with
position are explored in the Appendix.

The assumption that the distribution function depends only on energy is of
course not correct: even if it is for the initial model, anisotropy will
develop as the evolution proceeds.  The Fokker-Planck equation can be solved
with a distribution function that depends on both energy and angular
momentum, but the calculations are then much more difficult.  Calculations
done in this way for the collapse of an isotropic Plummer model show that
the velocity distribution remains nearly isotropic in the core (Cohn 1985,
Takahashi 1995).  A radial anisotropy develops in the outer parts of the
cluster. This is important for some aspects of the evolution, especially for
the escape of stars, but it does not cause a big change in the collapse
time.  Takahashi's calculation shows the late, self-similar stage of core
collapse to be about 40\% slower when anisotropy is included; earlier
calculations had suggested that it was 40\% faster.

The other assumptions made in deriving equation~(\ref{eq-fp})---the orbit
averaging, the local approximation, the neglect of close encounters---are
difficult to justify (Goodman 1983 modified the equation to include close
encounters and found that it did not change the evolution much).  Yet the
predictions for core collapse made with these assumptions agree well with
the results from large N-body experiments (Spurzem and Aarseth~1996).
Surprisingly, even experiments with small values of $N$ follow the
predictions when many experiments are averaged to reduce the noise (Giersz
and Heggie 1994).  There are some differences but the agreement is
impressive.

\subsection{Numerical solution}

The Fokker-Planck program of Quinlan and Shapiro (1989) was modified for
these calculations so that it could use models with density cusps. It is
based on the algorithm developed by Cohn~(1980).  Since the goal was to
compute accurate collapse times, the time steps, grid spacings, and
tolerance parameters were chosen smaller than is necessary for most
applications.  Those choices are described here along with some test
results.

The radial grid points were spaced equally in $\log(r)$ between inner and
outer boundaries $\rmin$ and $\rmax$ ($\rmin\simeq10^{-6}$ and $\rmax\simeq
10^{3}$; the exact numbers depend on the model).  Most calculations used
$\Delta\log(r)=0.05$; some used a smaller spacing for higher accuracy.  For
initial models with cores the energy grid points were chosen as in Cohn's
program; for models with cusps they were chosen to match the potential at
the radial grid points (the number of radial and energy grid points was the
same).  The equation was advanced with Crank-Nicholson time differencing and
Chang-Cooper space differencing; zero-flux boundary conditions were imposed
at the inner and outer grid points. The time step for the potential
recomputation was chosen so that the central density changed by at most one
percent between time steps; 32 Fokker-Planck steps were taken for each
potential step.  The recomputation was iterated until the potential
converged to one part in $10^5$ at all grid points (which usually took 8--10
iterations). The phase-space integrals $p$ and $q$ were computed to an
accuracy of about one part in $10^5$.

The ability of the program to reproduce the density and potential of the
initial models was checked; it did this to at least one part in $10^4$ at
all grid points (except for the last few near the outer boundary).  The main
test was the standard Plummer-model collapse.  This was followed until the
central density had risen by a factor of $10^7$, using 140 grid points with
$\rmin=5\times 10^{-5}$ and $\rmax=500$. Energy was conserved to one part in
$10^3$ and mass to one part in $10^5$. The collapse time was $15.4\,\trh$.
The results for the evolution of the collapse rate, $\xi=\trc\,\d\ln\rhoc/\d
t$, the equation of state, $\d\ln\vc^2/\d\ln\rhoc$, and the scaled central
potential, $x_0=3|\phic|/\vc^2$, agreed well with those published by
Cohn~(1980).

\section{Results}

\subsection{Notation}

Some definitions and conventions used throughout are gathered here for
convenience.  The local and half-mass relaxation times are computed from the
standard definitions (Spitzer 1987)
\begin{eqnarray}                                                  \label{eq-tr}
     \tr  &=& {0.065 v^3 \over G^2m\rho\ln\Lambda}, \\
                                                                 \label{eq-trh}
     \trh &=& {0.138 N\over \ln\Lambda}\left(\rh^3\over GM\right)^{1/2},
\end{eqnarray}
where $M=Nm$ is the total mass, $\rho$ the density, $v$ the
three-dimensional velocity dispersion, $\rh$ the half-mass radius, and
$\ln\Lambda$ the Coulomb logarithm.  The half-mass relaxation time does not
change by much during the evolution of the models studied here; $\trh$ will
always denote its initial value.  The central relaxation time $\trc$ is
given by equation~(\ref{eq-tr}) with $\rho$ and $v$ replaced by $\rhoc$ and
$\vc$. The subscript ``c'' indicates that a variable is to be evaluated at
the center; one exception is $\rc$ which denotes the core radius (sometimes
called the King radius)
\begin{equation}                                                  \label{eq-rc}
  \rc = \left(3\vc^2\over 4\pi G\rhoc\right)^{1/2}.
\end{equation}
If the units of measurement are not explained when results are presented in
the figures or text then they are the standard N-body units in which
$G=M=-4E=1$ (Heggie and Mathieu 1986).

\subsection{Models with cores}

\subsubsection{Initial models}

\begin{figure}[tb]                                              
\centerline{\psfig{figure=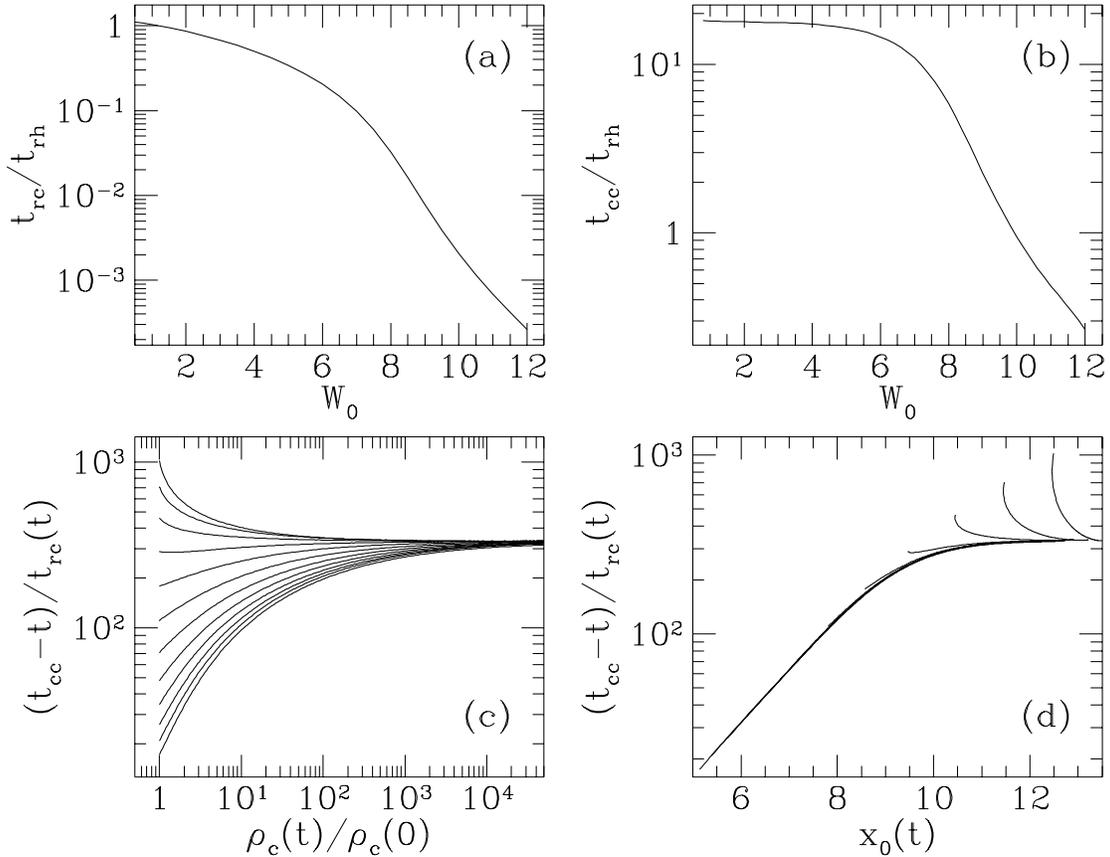,width=0.90\textwidth,clip=}}
\caption[Collapse of isolated King models.]{Collapse of isolated King
models: (a) ratio of initial central and half-mass relaxation times; (b)
collapse time; (c) time remaining until complete collapse during the
evolution of twelve models ($W_0=$1, 2, \ldots, 12, increasing from bottom
to top); (d) the same as in (c) but plotted versus $x_0=3|\phic|/\vc^2$.}
\label{fig-king} 
\end{figure}

King models (King 1966) provide a convenient one-parameter family of models
with cores.  They are lowered isothermals with two length scales: a tidal
radius $\rt$, at which the density drops to zero, and a core radius $\rk$
(which differs slightly from $\rc$ because its definition uses $3\sigma^2$
instead of $\vc^2$).  The concentration is measured by either the parameter
$c=\log(\rt/\rk)$ or the dimensionless central potential $W_0$, which will
be used here; the relation between the two is plotted in Figure~4.10 of
Binney and Tremaine~(1987).  The ratio of the central and half-mass
relaxation times varies by nearly a factor of $10^4$ over the range
$W_0=1$--12 (see Fig.~\ref{fig-king}).  In the limit of infinite
concentration the King models approach the isothermal sphere.

Two Fokker-Planck studies have followed the collapse of King
models: Wiyanto, Kato, and Inagaki~(1985) used equal-mass models with
$W_0=0.5$, 6.6, and 12.2; Chernoff and Weinberg (1990) used models with
$W_0=1$, 3, and 7, including a distribution of masses and stellar
evolution.  Both these studies assumed the models to be tidally truncated,
meaning that stars were removed if they moved outside the tidal radius and
that the tidal radius was reduced along with the mass as $\rt\sim M^{1/3}$.
In contrast to these, the present study considers a wide range of
concentrations but ignores mass loss: the models are isolated and stars
remain bound if they move outside the tidal radius.  Mass loss from a tidal
boundary reduces the collapse time (because it lowers $N$ while keeping the
mean density fixed, which lowers the relaxation time), but this is important
only for low-concentration models.

The Fokker-Planck program used here required the density to be non-zero at
all grid points. The King models were therefore modified at grid points
$r\geq\rt$ so that the density fell to a small value at $\rt$ and then
continued to fall with radius out to infinity.  The mass added in this way
was tiny and did not affect the collapse times.

\subsubsection{Evolution}

Each model was integrated until its core radius $\rc$ had shrunk to
$10^{-5}$; by then the central density is much higher and the evolution
time-scale much smaller than at the start.  An extrapolation was made to
predict the time $\tcc$ at which $\rc$ would reach zero. These collapse
times are plotted in panel~(b) of Figure~\ref{fig-king} in units of the
half-mass relaxation time (see also Table~\ref{tab-king}). For
low-concentration models ($W_0\leq 7$) they agree with the often-quoted
collapse time of 12--$19\,\trh$; for high-concentration models they are
shorter.

\begin{table}
\begin{center} 
\begin{tabular}{|r|r|r|r|r|}
\hline
$W_0$ & $c$ & $\trc/\trh$ & $\tcc/\trc$ & $\tcc/\trh$ \\
\hline
  1.0  &  0.296  &  1.032123  &   18.  &  18.12 \\
  2.0  &  0.505  &  0.859156  &   21.  &  17.94 \\
  3.0  &  0.672  &  0.678293  &   26.  &  17.70 \\
  4.0  &  0.840  &  0.502196  &   34.  &  17.33 \\
  5.0  &  1.029  &  0.341928  &   48.  &  16.45 \\
  6.0  &  1.255  &  0.205136  &   71.  &  14.56 \\
  7.0  &  1.528  &  0.098544  &  110.  &  10.90 \\
  8.0  &  1.833  &  0.032641  &  179.  &   5.84 \\
  9.0  &  2.118  &  0.007875  &  289.  &   2.28 \\
 10.0  &  2.350  &  0.002066  &  459.  &   0.95 \\
 11.0  &  2.548  &  0.000683  &  707.  &   0.48 \\
 12.0  &  2.739  &  0.000261  & 1018.  &   0.27 \\
\hline
\end{tabular}
\end{center} 
\caption{Collapse times for isolated King models (accurate to about one
percent).}
\label{tab-king}
\end{table}

Panel (c) plots for twelve models the time remaining until complete collapse
at any instant during the evolution, $\tau=\tcc-t$, in units of the central
relaxation time at that instant.  The abscissa is chosen to separate the
models; it obscures the fact that the King models form an evolutionary
sequence (King~1966, Cohn~1980, Wiyanto et al.~1985).  This is clearer in
panel~(d) where the times are plotted versus the scaled central potential
$x_0$: the lines for $W_0=1$--7 fall on top of each other, showing that
low-concentration King models evolve through states resembling models with
higher concentrations; around $W_0\simeq7$--8 the lines depart from this
evolutionary sequence, near the value $W_0=7.4$ at which King models become
unstable to the gravothermal catastrophe (Katz 1980).  Deep into the
collapse the values of $\tau/\trc$ for all the models converge to
$\tau/\trc\simeq330$, the same value found by Cohn (1980) for the late,
self-similar stage of the collapse of a Plummer model. But most of the time
is spent near the start where $\tcc/\trc$ can be quite different from 330:
it is much smaller for low-concentration models, and larger for
high-concentration models---about 1000 for $W_0=12$ and 2500 for $W_0=15$
(this last value is not shown in the figure).

\begin{figure}[tb]                                               
\centerline{\psfig{figure=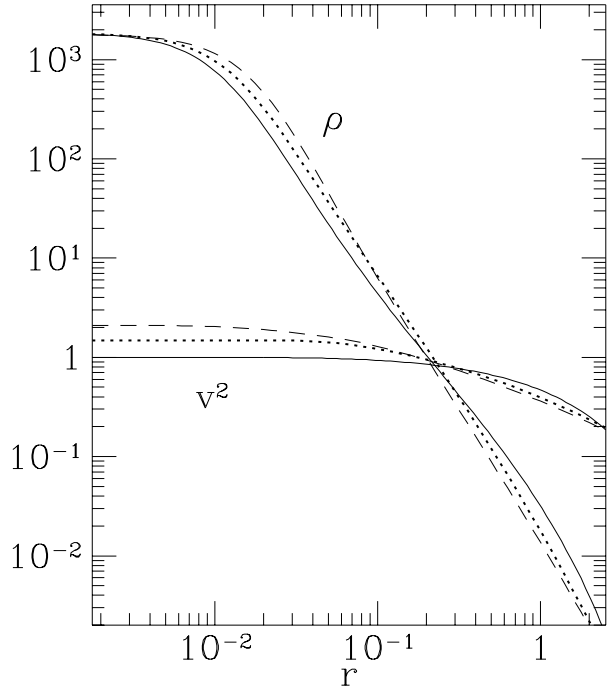,width=0.51\textwidth,clip=}}
\caption[High-concentration core models.]{Density and velocity dispersion
(squared) for the $W_0=12$ King model (solid lines) and for models with
$\rho\sim r^{-2.25}$ (dotted) and $\rho\sim r^{-2.5}$ (dashed).}
\label{fig-W12}
\end{figure}

The high-concentration King models have long collapse times (when measured
in units of $\trc$) because they are nearly isothermal; other
high-concentration models with cores have shorter collapse times.  Consider
the two models in Figure~\ref{fig-W12}.  They have densities $\rho(r) \sim
(r^2+b^2)^{-\gamma}(r+a)^{4-\gamma}$, with $a$ and $b$ chosen to give a
concentration like that of the $W_0=12$ King model (this is the
$\gamma$~model density from the next section, modified to have a core of
radius $\rc\simeq b$).  The model with $\gamma=2.25$ collapses in about 400
central relaxation times; the model with $\gamma=2.5$ in about 100, ten
times faster than the $W_0=12$ King model.

The amount by which mass loss from a tidal boundary reduces the collapse
time can be judged from the results of Chernoff and Weinberg~(1990).  They
give the collapse time for an equal-mass model without stellar evolution for
only one concentration ($\tcc =10.1\,\trh$ for $W_0=7$); times for two other
concentrations were kindly provided by M.~Weinberg ($9.6\,\trh$ for $W_0=3$
and $2.23\,\trh$ for $W_0=9$). The ratios of their times and those found
here are 0.54, 0.93, and 0.99 for $W_0=3$, 7, and 9.  Thus mass loss can
reduce the collapse time by about a factor of two for a low-concentration
model like the $W_0=3$ King model (which loses 75\% of its mass during the
collapse), but it does not reduce the time by much for high-concentration
models.\footnote{The collapse times of Wiyanto et al.\ (1985) are
questionable (note that their definition of $\trc$ differs from the
definition used here).  Their time for $W_0=0.5$ is about three times
shorter than found here, which seems reasonable, but their times for
$W_0=6.6$ and 12.2 are longer than found here, by factors of 1.3 (for
$W_0=6.6$) and 2.6 (for $W_0=12.2$).  They do not describe the accuracy of
their integrations.  The integrations done here were repeated with finer
grid spacings and smaller time steps and tolerance parameters to check that
the times had converged.  This required higher accuracies for the
high-concentration models than for the low-concentration models.}

\subsection{Models with cusps}

\subsubsection{Initial models}

A simple one-parameter family of models with density cusps is provided by
the density (Carollo 1993, Dehnen 1993, Tremaine et al.\ 1994)
\begin{equation}                                               \label{eq-gamma}
   \rho(r) = {3-\gamma\over 4\pi}{Ma\over r^\gamma (r+a)^{4-\gamma} }.
\end{equation}
In standard N-body units the length scale is $a=1/(5-2\gamma)$.  These will
be called $\gamma$~models; they include as special cases the models of
Hernquist (1990), $\gamma=1$, and Jaffe (1983), $\gamma=2$.  For most
$\gamma$~values the distribution function must be computed numerically from
Eddington's formula.  The $\gamma$~models are themselves special cases of a
more general family considered by Zhao (1996),
\begin{equation}                                                \label{eq-zhao}
   \rho(r) = { C \over r^\gamma (r^{1/\alpha}+a)^{(\beta-\gamma)\alpha} },
\end{equation}
which resembles the fitting formula used by Lauer et al.\ (1995) and Byun et
al.\ (1996) for the central regions of elliptical galaxies (but they fit the
surface brightness, not the density).  The extra flexibility provided by
$\alpha$ and $\beta$ is not needed here because the goal of the calculations
is to understand how relaxation affects the central region; that is
determined primarily by $\gamma$.

The velocity dispersion near the center of a $\gamma$~model varies with
radius as $v^2\sim r^\gamma$ if $\gamma<1$ and as $r^{2-\gamma}$ if
$\gamma>1$.  The models with $\gamma<2$ have a temperature inversion: the
maximum of $v^2$ is not at the center (``temperature'' and ``$v^2$'' can be
used synonymously when the stars all have the same mass).  This property is
shared by all isotropic models with density cusps more gradual than $r^{-2}$
(Binney 1980).  Relaxation tends to even out the temperature in a star
cluster; energy therefore flows into the center when there is an inversion,
causing the central region to expand.  This is familiar from Fokker-Planck
studies of the post-collapse evolution of globular clusters: the collapse is
stopped by something that heats the core (usually binaries), either directly
(by increasing the kinetic energy of the stars) or indirectly (by ejecting
mass from the core); once the core expands a little it becomes cooler than
the surrounding medium and can continue expanding even if the heat source is
turned off---the expansion is then said to be gravothermal.  Heggie,
Inagaki, and McMillan~(1994) verified that an N-body system expands when it
starts with an inversion like that expected for globular clusters after core
collapse; the larger N-body experiments of Makino (1996) show that binaries
can produce the required inversion and can cause alternating periods of
contraction and expansion known as gravothermal oscillations.  The
$\gamma$~models do not need a heat source to start the expansion because
they start with a temperature inversion.

\begin{figure}[tb]                                               
\centerline{\psfig{figure=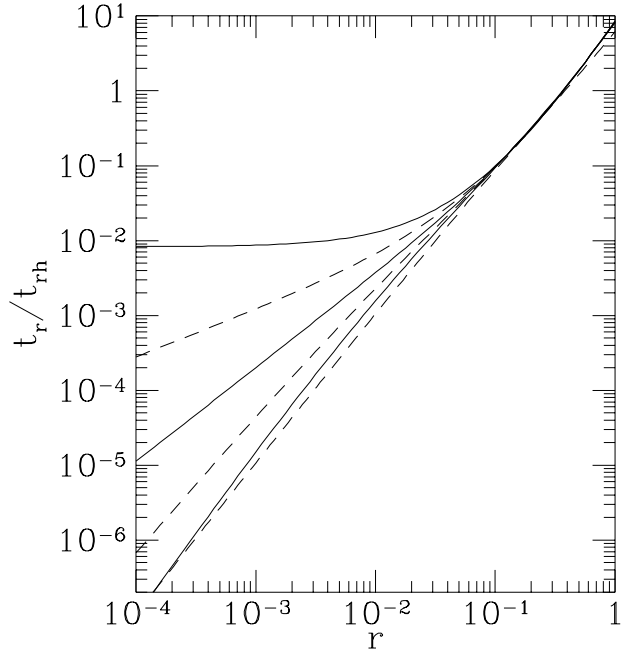,width=0.51\textwidth,clip=}}
\caption[Relaxation time in a $\gamma$~model.]{
The local relaxation time in $\gamma$~models with $\gamma = 0.0$ (top line,
solid), 0.25 (dashed), 0.5 (solid), 0.75 (dashed), 1.0 (solid), and 2.0
(bottom line, dashed).}
\label{fig-gtr}
\end{figure}

The rising density and (for $\gamma<2$) falling velocity dispersion near the
center of a $\gamma$~model cause a rapid fall in the relaxation time (see
Fig.~\ref{fig-gtr}), with $\tr\sim r^{5\gamma/2}$ for $\gamma<1$ and $\sim
r^{3-\gamma/2}$ for $\gamma>1$ (the fastest fall occurs for the Hernquist
model).  This causes obvious problems for the Fokker-Planck program: the
evolution is arbitrarily fast near the center and the zero-flux boundary
condition cannot be satisfied.  The problems can be avoided by replacing the
$r^{-\gamma}$ in the density law by $(r^2+b^2)^{-\gamma/2}$ to give the
models a small core (the experiments described below used $b=10^{-5}$).  The
models with $\gamma<2$ would develop cores in any case once the expansion
begins, so it does no harm to give them one at the start provided that $b$
is smaller than the length scales of interest.  The models with
$\gamma\geq2$ are not considered here because they do not have temperature
inversions and do not expand; if given a core of size $b$ they collapse in a
time determined by $b$, as happened for the two models in
Figure~\ref{fig-W12}.

\subsubsection{Evolution}

\begin{figure}[tb]
\centerline{\psfig{figure=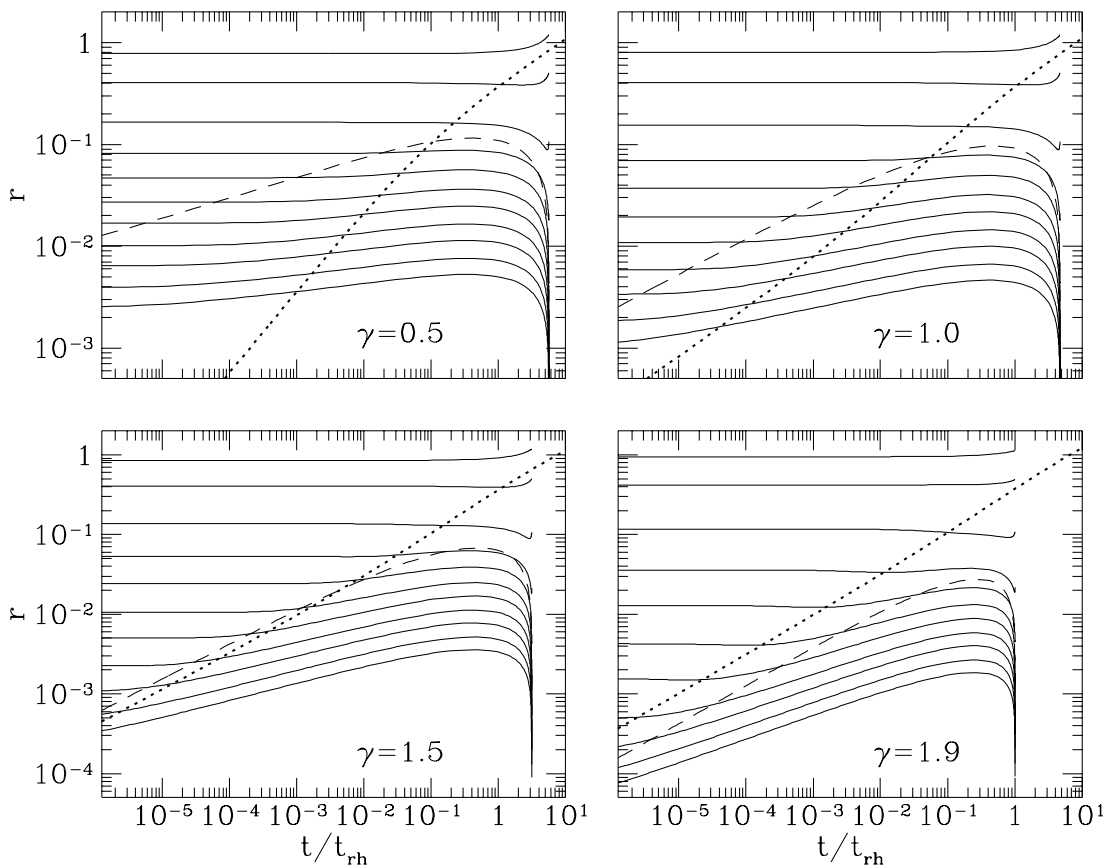,width=0.90\textwidth,clip=}}
\caption[Expansion of a gamma models.]{Expansion and collapse of
$\gamma$~models. The solid lines are radii containing fixed fractions of 
the total mass ($10^{-5}$, $3\times10^{-5}$, $10^{-4}$, $3\times10^{-4}$,
\ldots, 0.1, 0.3, 0.5); the dashed line is the core radius.  The dotted line
shows the local relaxation time versus radius in the initial model.}
\label{fig-lagr}
\end{figure}

\begin{figure}[tb]
\centerline{\psfig{figure=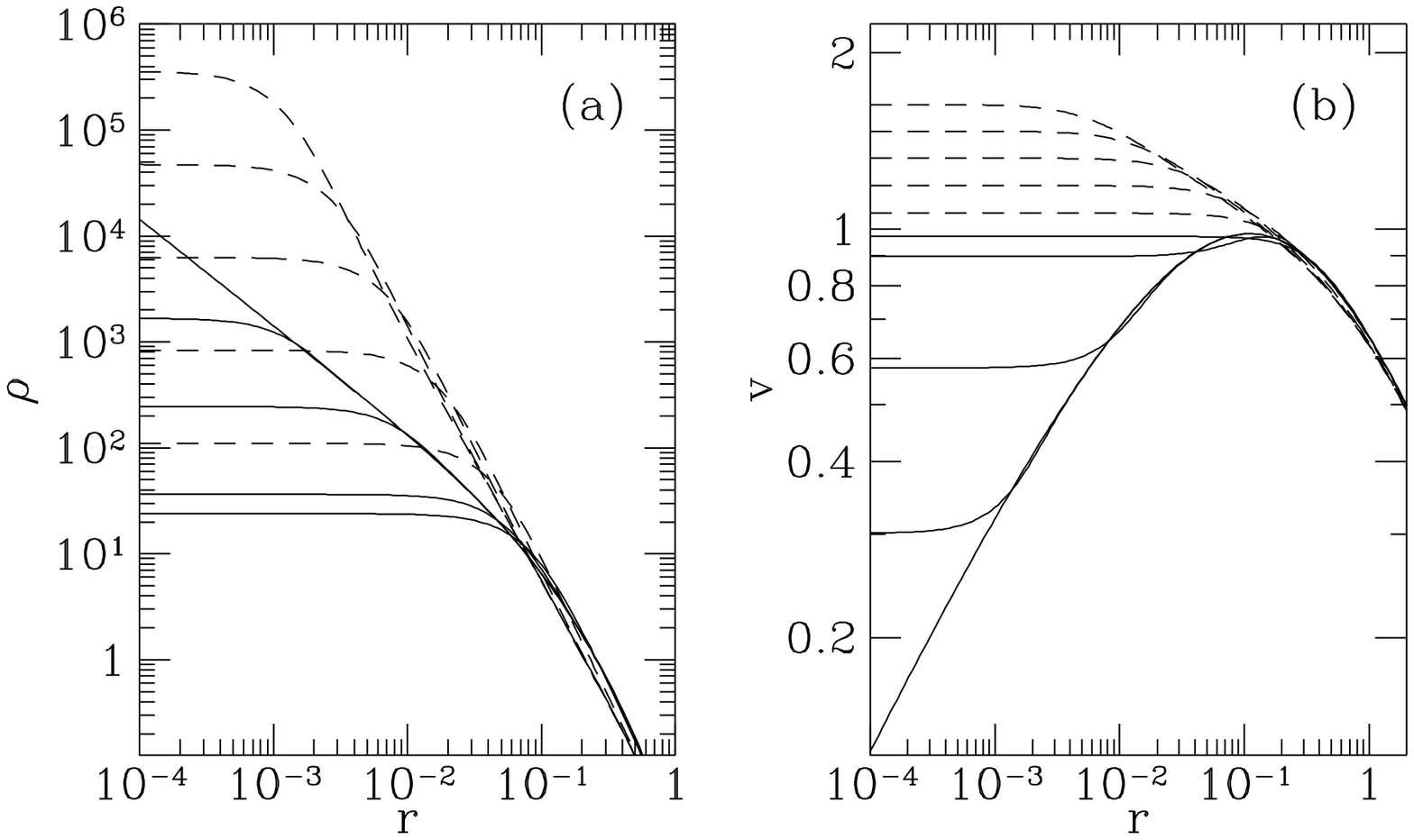,width=0.90\textwidth,clip=}}
\caption[Core-collapse of a Hernquist model.]{Density and velocity
dispersion at ten times during the expansion (solid lines) and collapse
(dashed lines) of a Hernquist model ($\gamma=1$).  The first solid line (the
highest for $\rho$, the lowest for $v$) shows the initial model; the last
shows the point of maximum expansion.}
\label{fig-hern}
\end{figure}

The results from the Fokker-Planck calculations show that the models with
$\gamma<2$ expand gravothermally, as expected (see Figs.~\ref{fig-lagr} and
\ref{fig-hern}).  They develop cores that grow outward, with the central
density falling and the velocity rising (the outer parts contract slightly
to conserve energy).  The expansion continues until the temperature
inversion is gone; then it reverses and the core collapses.  The time to
reach complete collapse from the start gets shorter as $\gamma$ approaches 2
(and the extent of the expansion gets smaller), but it is always long if
$\trh$ is long.

\begin{figure}[tb]
\centerline{\psfig{figure=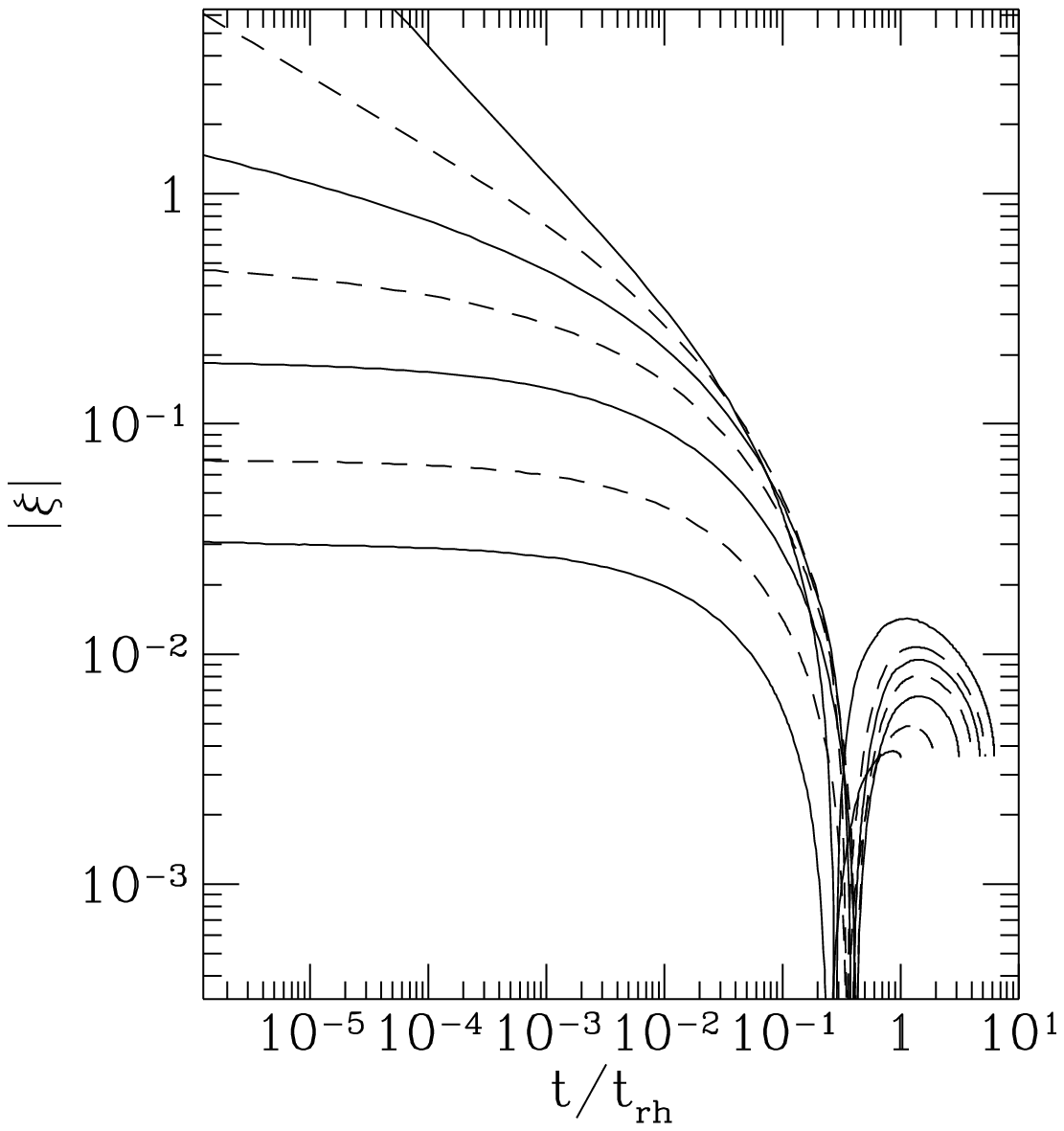,width=0.51\textwidth,clip=}}
\caption[Expansion of a gamma models.]{Evolution of $\xi = \trc\,
\d\ln\rhoc/\d t$ for models with $\gamma=0.0$ (solid, top line), 0.75
(dashed), 1.0 (solid), 1.25 (dashed), 1.5 (solid), 1.75 (dashed), and 1.9
(solid, bottom line). $\xi$ is negative at first, then changes to positive.}
\label{fig-xi}
\end{figure}

The expansion time is more interesting than the collapse time for these
models.  In the first three panels of Figure~\ref{fig-lagr} (the models with
$\gamma\leq1.5$) the solid lines near the center rise noticeably before they
cross the dotted line, showing that the local expansion time is no longer
than the relaxation time, unlike in globular clusters where the expansion
time after core collapse is usually a thousand or more relaxation times
(e.g.\ Heggie and Ramamani~1989).  Another measure of the expansion rate is
plotted in Figure~\ref{fig-xi}: $|\xi|$ is of order unity for the
$\gamma$~models with $\gamma\simless 1$, much larger than the values of
$|\xi|\simless10^{-3}$ typical for globular clusters undergoing gravothermal
oscillations (see Fig.~2 of Cohn, Hut, and Wise 1989).  The reason the
$\gamma$~models expand so fast is that they start far from thermal
equilibrium.  As $\gamma$ approaches 2 the initial temperature inversion
gets weaker and the expansion gets slower.

Other isotropic models with power-law density cusps will expand in about the
same way as the $\gamma$~models. The Zhao models with $0<\gamma<2$, for
example, all have temperature inversions; the parameters $\alpha$ and
$\beta$ will modify the extent of the expansion but not its initial speed.

\subsection{Models with nuclei}

King models and $\gamma$~models do not exhaust the possibilities for
spherical star clusters.  Some galaxies have density cusps that steepen near
the center; others have dense cores embedded in larger cores or weak cusps.
Lauer et al.~(1995) describe these as galaxies with nuclei.  Their evolution
cannot be surveyed with a simple one-parameter family of models, but it can
be described easily in two limits.  If the nucleus has a short relaxation
time, it collapses or expands as it would in isolation; if it consists of
low-mass stars with a long relaxation time, it traps the more massive stars
from the rest of the galaxy and causes them to collapse.  Examples of both
limits are described below.

\subsubsection{Equal-mass models}

\begin{figure}[tb]                                              
\centerline{\psfig{figure=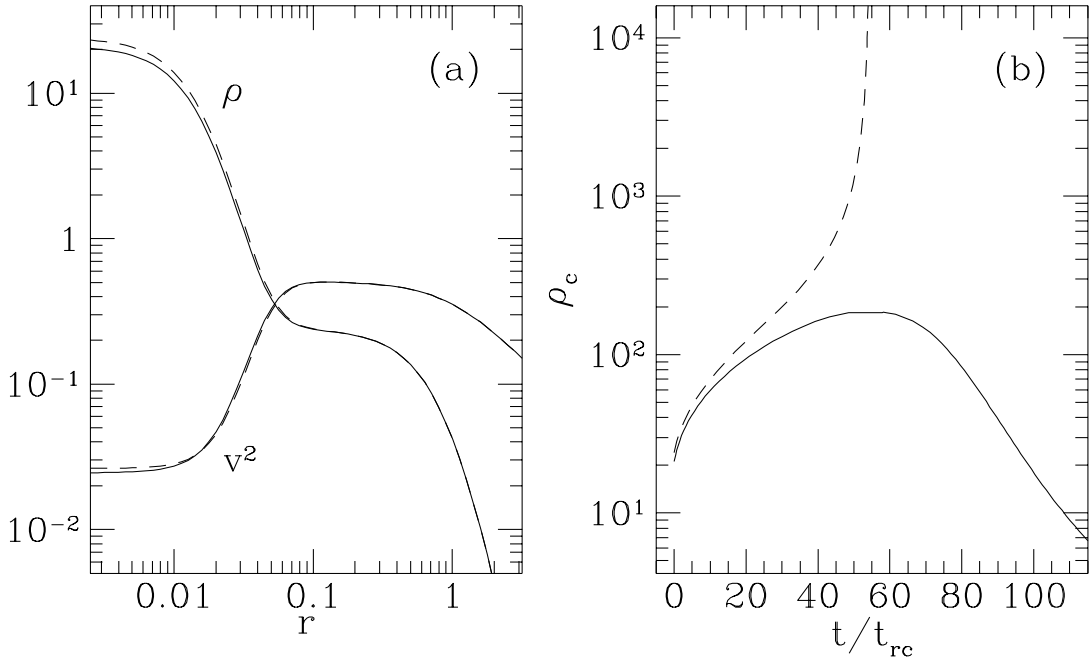,width=0.90\textwidth,clip=}}
\caption[Core-collapse of a double-Plummer model.]{Models with nuclei
formed by combining the densities of Plummer models with radii $R_1=0.02$
and $R_2=1.0$ and masses $M_1$ and $M_2=1-M_1$: (a) initial density and
squared velocity dispersion for $M_1=7\times10^{-4}$ (solid lines) and
$8\times10^{-4}$ (dashed lines); (b) evolution of the central density for
the two models ($\trc$ is the initial central relaxation time).}
\label{fig-dbpl}
\end{figure}

A nucleus with a short relaxation time can collapse even if its temperature
is lower than the temperature outside the nucleus.  As a simple example,
consider a model whose density is the sum of two Plummer-model densities, an
inner model with mass $M_1$ and radius $R_1$ and an outer model with mass
$M_2$ and radius $R_2>R_1$ (the Plummer potential is $\phi =
-GM/\sqrt{r^2+R^2}$).  If $M_1/R_1^3>M_2/R_2^3$ the inner model has a higher
density than the outer model and appears as a nucleus.  Yet if
$M_1/R_1<M_2/R_2$ the inner model has a lower temperature than the outer
model; whether it collapses or expands then depends on whether its collapse
time is shorter or longer than the time for it to absorb energy from the
outer model. Figure~\ref{fig-dbpl} shows models on the two sides of this
boundary: the one with $M_1=8\times10^{-4}$ collapses despite its
temperature inversion; the one with $M_1=7\times10^{-4}$ collapses at first
but then stops and expands.  Experiments like this were done for models with
$0.005\leq R_1/R_2 \leq 0.08$; the boundary between the collapsing and
expanding models is fit well by $M_1/M_2 = 1020 (R_1/R_2)^{3.6}$ over this
range.  If $M_1$ is just a few times larger than the boundary then the
collapse time is about the same as for an isolated Plummer model
($33\,\trc$).

\subsubsection{Two-component models}

The collapse time for a nuclear star cluster can be lengthened by lowering
the mass $m_1$ of its stars, because $\trc\sim 1/m_1$ when the density is
held fixed.  But the more massive stars from the rest of the galaxy that
pass through the nucleus then get trapped by dynamical friction in a time
that is independent of $m_1$.  This is of interest for models of galaxies
with central mass concentrations, for which dark clusters of low-mass stars
(which might be brown dwarfs or planets or small black holes) are sometimes
proposed as alternatives to massive black holes.

\begin{figure}[tb]
\centerline{\psfig{figure=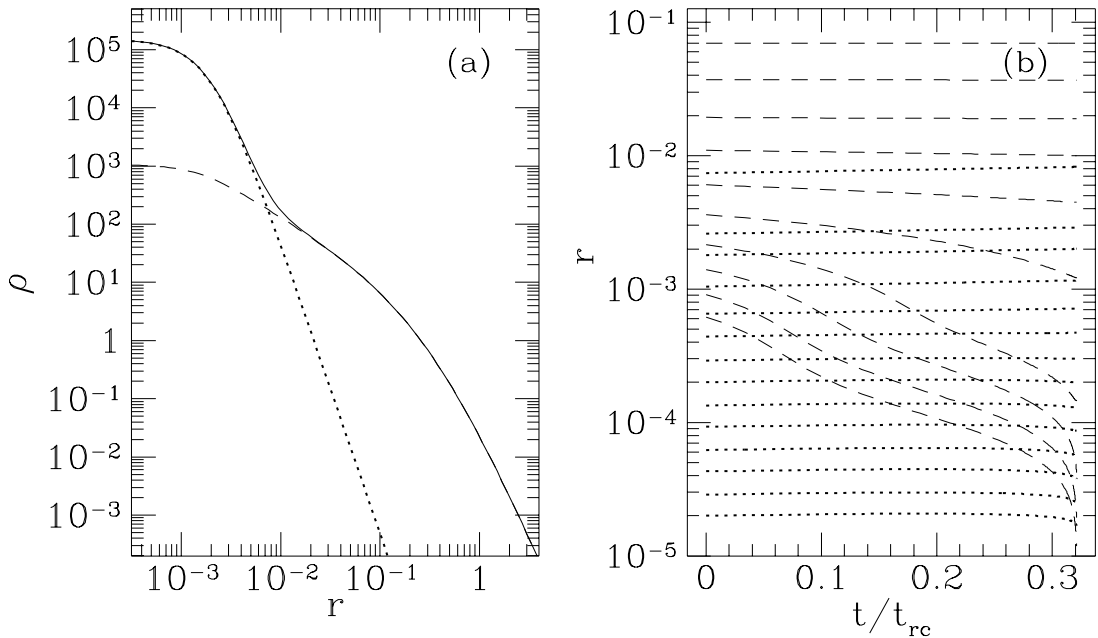,width=0.90\textwidth,clip=}}
\caption[Core-collapse of a gamma-Plummer model.]{(a) A model formed by
combining the densities of a Plummer model with $M_1=0.005$ and $R=0.002$
(dotted line) and a $\gamma$~model with $\gamma=1$, $M_2=0.995$, $a=1/3$,
and $b=0.002$ (dashed line).  (b) The evolution when $m_2/m_1=100$: the
dotted lines show radii containing fixed fractions of the mass of the first
component ($10^{-6}$, $3\times 10^{-6}$, $10^{-5}$, $3\times 10^{-5}$,
\ldots, 0.1, 0.3, 0.5, 0.9); the dashed lines show the same for the second
component (the larger fractions are not visible).}
\label{fig-dgpl}
\end{figure}

The simple model used in Figure~\ref{fig-dbpl} cannot be used to illustrate
this limit because its distribution function cannot be split into positive
functions $f_1$ and $f_2$ that generate the densities of the two Plummer
models, for the same reason that Tremaine et al.~(1994) could not find
positive distribution functions for $\gamma$~models with central black holes
when $\gamma<1/2$.  Instead we shall consider the more realistic model in
Figure~\ref{fig-dgpl} that has a Plummer-model nucleus embedded in a
$\gamma$~model with $\gamma=1$.  If the two components have equal stellar
masses ($m_1=m_2$) the nucleus collapses independently of the $\gamma$
model, in about the same time as for an isolated Plummer model.  Panel~(b)
shows what happens when the stellar mass is 100 times smaller in the nuclear
component than in the $\gamma$-model component.  The inner parts of the
$\gamma$ model sink to the center and collapse in a time of about
$0.3\,\trc$, 100 times shorter than the collapse time for the nucleus if it
were isolated.

Experiments with other values of $M_1$, $R$, and $\gamma$ gave similar
results when $m_2/m_1\gg1$.  The stars of the $\gamma$ model that start
within the nucleus collapse in a time that is $m_2/m_1$ times shorter than
the collapse time for the nucleus if it were isolated.  Additional stars
will be captured on longer time scales as they get scattered into loss-cone
orbits that bring them into the nucleus.  The mass that collapses could be
reduced by removing stars from the center of the $\gamma$ model, which would
require an anisotropic distribution function with a circular bias near the
center, but such a model would be contrived.

\section{Applications}

\subsection{Globular clusters}

The importance of relaxation for globular clusters is clear. Most Galactic
clusters are fit well by King models---models constructed for relaxed,
tidally-truncated star clusters.  The majority have concentrations of
$W_0\simeq 6$--8; fewer than 10\% have $W_0<4$; about 20\% have high
concentrations ($W_0>10$) or are not fit well by any concentration---these
are the ``core-collapsed'' or ``post-core-collapsed'' clusters (Trager,
King, and Djorgovski 1995).  Their fraction is consistent with simple
estimates of the collapse rate based on an assumed collapse time that is a
fixed multiple of the central relaxation time (Cohn and Hut 1984, Djorgovski
and Hut 1992).

\begin{figure}[tb]
\centerline{\psfig{figure=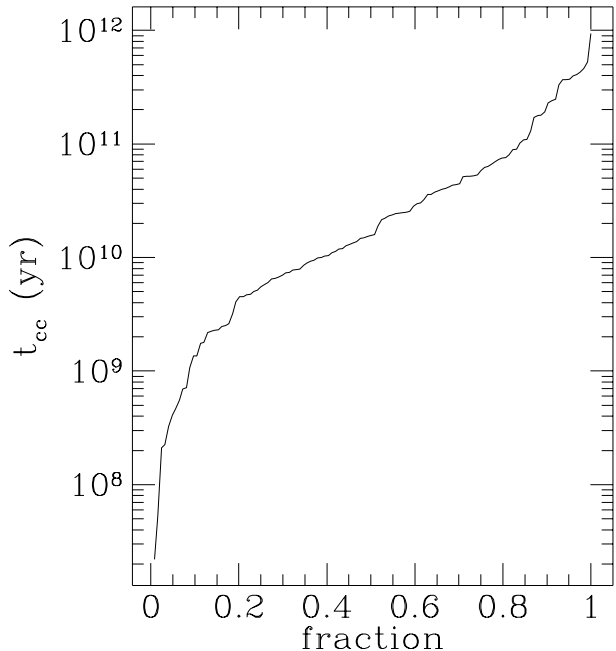,width=0.51\textwidth,clip=}}
\caption[Collapse times for globular clusters.]{Fraction of Galactic
globular clusters with collapse times smaller than $\tcc$.}
\label{fig-gc}
\end{figure}

In reality the $\tcc/\trc$ ratio depends on the concentration of a cluster;
this can be taken into account.  Djorgovski~(1993) and Trager et al.~(1995)
give values of $\trc$ and $W_0$ (or $c$) for 124
clusters\footnote{Djorgovski's $\trh$ values are a factor of $\ln(10)$ too
small because of an error in his equation~(11) (Djorgovski, private
communication). The data are available from a catalog maintained by
W.~E.~Harris (at http://www.physics.mcmaster.ca/Globular.html).}; these were
combined with the data from Figure~\ref{fig-king} to predict the time $\tcc$
for each cluster to collapse from its present state, assuming that the
clusters are collapsing and that they are isolated, equal-mass King models.
Figure~\ref{fig-gc} shows the resulting distribution.  About 20\% of the
clusters have short collapse times ($\tcc\simless 4\times10^9\,\yr$),
consistent with the suggestion that many of these have already passed
through core collapse (Djorgovski and King~1986).

The collapse times computed here ignore several complicating factors known
to be important for real clusters (reviewed by Chernoff 1993), including
stellar evolution, mass segregation, and mass loss from a tidal boundary.
Stellar evolution always slows the collapse, and can cause some clusters to
disrupt; its importance depends on the initial mass function.  Mass
segregation always accelerates the collapse, by as much as a factor of five
or ten for a Salpeter-like mass function (Murphy and Cohn~1988, Chernoff and
Weinberg~1990).  Mass loss from a tidal boundary is more complicated: if the
tidal field is constant, the mass loss accelerates the collapse of
low-concentration clusters, by a factor of two or three for $W_0\simless3$;
but if the mass loss is enhanced by gravitational shocks as the clusters
pass through the disk or close to the center of the Galaxy, it tends to
disrupt low-concentration clusters, though it accelerates the collapse of
others.  The increase in cluster concentrations towards the center of the
Galaxy shows the importance of these environmental factors (Chernoff and
Djorgovski~1989, Djorgovski and Meylan~1994).  The clusters would have to be
modelled on a case by case basis to take all the complications into account;
but since they tend mostly to accelerate the collapse it seems that the
fraction of clusters with short collapse times is probably larger than shown
in Figure~\ref{fig-gc}.

\subsection{Galaxies}

The relaxation time is known to be short in the centers of several
local-group galaxies: about $2\times 10^{7}\yr$ in the nucleus of M33
(Kormendy and McClure 1993); $10^{8}\yr$ in the inner $0.5\pc$ of our Galaxy
(Genzel, Hollenback, and Townes 1994); $4\times 10^9\yr$ or $6\times
10^7\yr$ in the inner $1\pc$ of M32, for models with and without massive
black holes (Lauer et al.\ 1992).  And short times like these are likely to
be common: many galaxies have density cusps that rise as steeply as in our
own Galaxy and M32 (Crane et al.\ 1993, Ferrarese et al.\ 1994, Forbes et
al.\ 1995, Lauer et al.\ 1995).  If the galaxies are like isotropic
$\gamma$~models (which here includes Zhao's more general class of models),
the ones with cusp slopes of $\gamma<2$ will be expanding. The cusps in
those models cannot survive for much longer than the local relaxation time.
For distant galaxies the region where the relaxation time is short cannot be
resolved, but for nearby galaxies the consequences of the expansion should
be observable.

But it is not clear that real galaxies are like isotropic $\gamma$~models.
The models fit the surface brightness of galaxies well but not the
kinematics.  Few galaxies show evidence of the temperature inversion that
the models predict.  The inversion is weakened when the velocity dispersion
is projected along the line of sight, but it should still be
observable\footnote{The figures of Tremaine et al.\ (1994) showing the
intrinsic and projected dispersions are interchanged: Figure~1 should be
Figure~4 and vice versa.}.  The elliptical galaxy NGC~5813 has a dispersion
that falls by about $30\,$km/s towards the center (Efstathiou, Ellis, and
Carter~1982), but it does not have an isotropic, power-law cusp: its inner
core rotates rapidly and appears to be kinematically distinct;
Kormendy~(1984) has argued that it is a merger remnant.  Five of the
forty-four elliptical galaxies in the survey of Bender, Saglia, and
Gerhard~(1994) have falling central dispersions, but two of these again have
rapidly rotating cores (Bender et al.\ note that $v^2+v_{\rm rot}^2$ remains
more constant with radius than $v^2$ does); the other thirty-nine have
constant or rising dispersions.  Byun et al.\ (1996) have fit isotropic,
power-law cusps to the central regions of fifty-seven elliptical galaxies;
the fits for about half predict falling dispersions, yet the data for nearly
all show constant or rising dispersions (Tremaine, private communication).

Although the discrepancy between the data and the fits is not large (about
10\% on average), it suggests that the galaxies differ from isotropic
$\gamma$~models.  It could be that the velocity distributions are not
isotropic; a radial bias can raise the dispersion at the center (Tonry
1983).  Even if they are isotropic, the dispersion need not fall if there is
a steepening density cusp (e.g.\ Binney 1982) or a rising mass-to-light
ratio or a central mass concentration such as a massive black hole.
Whatever the reason, relaxation will not change the galaxies as rapidly if
their velocity dispersions do not fall because their relaxation times will
be longer and their central regions will not be as far out of thermal
equilibrium.  In galaxies with central black holes the relaxation is
enhanced by resonances (Rauch and Tremaine~1996), but this affects only the
stars' angular momenta and hence does not lead directly to collapse or
expansion.

\subsection{Dark star clusters in the centers of galaxies}

There is growing evidence for dark mass concentrations in the centers of
galaxies, often called massive dark objects (Kormendy and Richstone~1995).
They are probably massive black holes, but to make that convincing we must
rule out alternatives such as clusters of dark stars (e.g.\ stellar
remnants, brown dwarfs, planets).  Various arguments can be made depending
on the constraints on the cluster and the assumed mass and size of its stars
(Goodman and Lee~1989, Maoz~1995).  Here we shall consider only one---that
the cluster is unacceptable if it will collapse in a small fraction of its
age.

The collapse time is of course unknown because the distribution of mass in
the cluster is unknown.  But since the goal is to argue that a dark cluster
is unacceptable, the collapse time should be chosen in the most generous way
to allow for its survival.  The best choice is therefore the time for a
low-concentration cluster of equal-mass stars, about $10\,\trh$ (where
$\trh$ refers to the dark cluster, not to the rest of the galaxy).
High-concentration clusters and clusters with a distribution of stellar
masses collapse faster than this; high-concentration clusters will suffer
from other problems too (as will clusters with density cusps)---their high
central densities will lead to stellar collisions and mergers.

Maoz~(1995) argued that the massive dark object in NGC~4258 cannot be a
cluster of objects of mass $\simgreat 0.03\,M_\odot$ because it would
collapse in $\simless 6\,$Gyr.  He used for the collapse time an evaporation
time of $136\,\trh$; his argument would have been stronger if he had used
the time of $10\,\trh$ suggested here (Maoz gave other arguments against a
dark cluster).  Goodman and Lee~(1989) did use a time of $10\,\trh$ in their
constraints for M31 and M32, but they assumed that the collapse had been
stopped and reversed (by binary heating or mass loss), and argued that
$10\,\trh$ could not be much less than the age of an expanding cluster.
This is a valid argument if the collapse can be stopped; whether that would
happen in M31 and M32 is debatable.  At the high velocity dispersions common
in galactic nuclei the collapse can continue all the way to the formation of
a massive black hole (Quinlan and Shapiro 1989, 1990).  Goodman and Lee's
constraints are applicable even if the collapse cannot be stopped, however,
because a dark cluster is unacceptable whether collapsing or expanding if
$10\,\trh$ is much less than its age.

The collapse time for a dark cluster can be made arbitrarily long by giving
its stars an arbitrarily small mass, and the problems with collisions and
mergers can be avoided by assuming the ``stars'' to be elementary particles
or small black holes. But then the cluster can capture stars from the rest
of the galaxy and cause them to collapse in a short time (as happened in the
model in Fig.~\ref{fig-dgpl}).  This gives another constraint on the size of
the cluster, because if too many stars get captured they might collapse to a
massive black hole, making the cluster an unacceptable alternative.

\section*{Acknowledgements}

I thank Martin Rees for suggesting the importance of mass segregation in
models like that shown in Figure~\ref{fig-dgpl}, and Douglas Heggie, Tad
Pryor, Scott Tremaine, and Martin Weinberg for helpful discussions about
other parts of the work.  Financial support was received from NSF grant ASC
93-18185 at UCSC and from NSF grant AST 93-18617 and NASA Theory grant NAG
5-2803 at Rutgers.

\appendix
\section{Calculations with variable Coulomb logarithm}

The perturbations to a star's energy from stars with different impact
parameters combine in such a way that, in a homogeneous cluster, equal
logarithmic intervals in impact parameter contribute equal amounts.  This is
the origin of the Coulomb logarithm,
\begin{equation}                                                \label{eq-clog}
  \ln\Lambda = \ln(\bmax/\bmin) = \ln(v^2\bmax/Gm),
\end{equation}
with $\bmin$ the impact parameter corresponding to a 90-degree deflection.
For low-concentration clusters $\bmax$ is usually taken to be the size of
the cluster (Farouki and Salpeter 1995); then $\Lambda=kN$ with $k\simeq0.4$
($k$ is used here instead of the usual $\gamma$ to avoid confusion with the
$\gamma$~models).  Giersz and Heggie (1994) found that $k=0.11$ works better
than $k=0.4$ when N-body experiments and Fokker-Planck and gas-model
calculations are compared for the collapse of a Plummer model, perhaps
because of the contribution from non-dominant terms in the perturbation
integrals (H\'enon~1975).  This difference will be ignored here.  For
high-concentration clusters it is not clear that any value of $k$ will work
well because equal logarithmic intervals in impact parameter do not
contribute equal amounts when the density varies with radius.  Binney and
Tremaine (1987, p.~511) suggest that for such clusters $\bmax$ should be the
radius of a star's orbit.  Giersz and Heggie~(1994) used a similar choice in
some of their gas-model calculations---they took $\Lambda(r) \sim
\mbox{Max}\{\Nc,N(r)\}$ and found that it slowed the collapse (they did not
say by how much).

The variation of the Coulomb logarithm with radius can be included in the
Fokker-Planck equation by including the logarithm in the integrals that
result from the orbit-averaging process.  The $p$ and $q$ integrals in the
definitions of $\De$ and $\Dee$ (eqns.~\ref{eq-de} and~\ref{eq-dee}) must be
replaced by
\begin{equation}                                                 \label{eq-pqb}
   \pl(E) = \deriv{\ql}{E}, \qquad
   \ql(E) = {1\over3}\int_0^{\phi^{-1}(E)}\!\! dr\,r^2
                   \left[2(E-\phi)\right]^{3/2} \ln\Lambda(r) ;
\end{equation}
the $p$ and $q$ in equation~(\ref{eq-fp}) are not changed.  In the
experiments described below $\Lambda(r)$ was taken to be
\begin{equation}                                               \label{eq-clogr}
   \Lambda(r) = {Nv^2(r)\over GM}\mbox{Max}\{r,\rc\}.
\end{equation}
For a Plummer model this gives $\Lambda\simeq0.5N$ in the halo and
$\Lambda\simeq0.35N$ in the core, close to the usual choice of $0.4N$.  The
value of $N$ must be specified at the start of the calculation; it cannot be
absorbed into the unit of time as it can when a constant logarithm is used.

\begin{figure}[tb]
\centerline{\psfig{figure=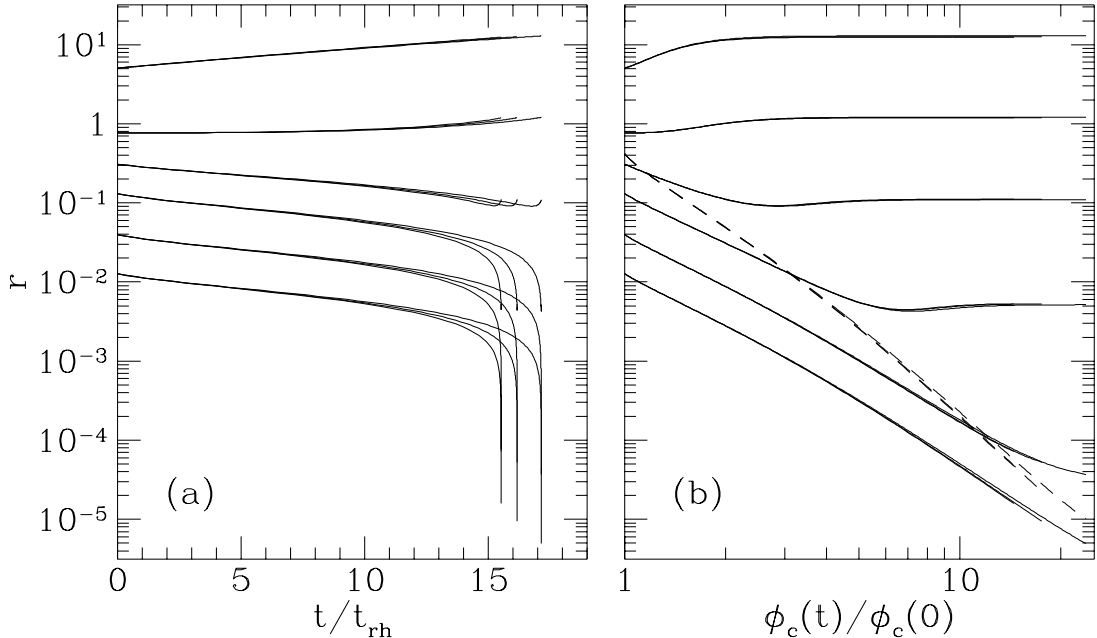,width=0.90\textwidth,clip=}}
\caption[Core-collapse with a variable Coulomb logarithm.]{Collapse of a
Plummer model with a Coulomb logarithm that depends upon radius. The solid
lines are radii containing fixed fractions of the total mass ($10^{-5}$,
$3\times10^{-4}$, $10^{-2}$, 0.1, 0.5, 0.9); the dashed line is the core
radius.  There are three sets of lines, for $N=10^4$ (the slowest collapse),
$10^8$, and $\infty$ (the fastest collapse).}
\label{fig-clog}
\end{figure}

Figure~\ref{fig-clog} compares the evolution of Plummer models computed
using two values of $N$ ($10^4$ and $10^8$) and using a constant Coulomb
logarithm (which corresponds to $N=\infty$).  Energy and mass were conserved
as well with the variable logarithm as with the constant logarithm.  The
main difference between the calculations is the collapse time, with
$\tcc/\trh$ equal to 17.1, 16.1, and 15.4 for $N=10^4$, $10^8$, and $\infty$
($\trh$ is computed from the usual definition with $k=0.4$; the $\tcc/\trh$
values change slightly if a different choice is made for $k$).  The
difference between the calculations is much smaller if the results are
compared at the same central potential instead of at the same time.  The
three sets of lines in panel~(b) lie almost on top of each other; the only
difference noticeable is that the core radius does not fall as rapidly with
the potential near the end of the calculation with $N=10^4$ as it does with
the other two $N$ values.  The collapse rate, $\xi=\trc\,\d\rhoc/\d t$, has
about the same value during the three calculations if the appropriate value
for $\Lambda$ in the core is used in the definition of $\trc$ (although near
the end $\xi$ is slightly larger with $N=10^4$ than with the other two $N$
values).

Calculations done with $\gamma$~models gave similar results. The evolution
is slower when the Coulomb logarithm varies with radius because the
relaxation time near the center is then longer, but the difference is large
only at radii where $\Lambda\simeq 1$.  For most time-scale estimates it can
be allowed for by evaluating $\Lambda$ at the radius where it is needed.
The one exception might be detailed comparisons of N-body experiments with
Fokker-Planck or gas-model calculations; small changes to the collapse time
can be important for that.  There are of course factors besides the Coulomb
logarithm that can cause such changes (Giersz and Heggie~1994, Giersz and
Spurzem~1994).


\vfill
\end{document}